\documentclass{article}
\usepackage{amsfonts}
\usepackage{amsmath}
\usepackage{graphicx}
\usepackage{xcolor}
\usepackage{overpic}
\usepackage{indentfirst}

\newcommand{\bea}{\begin{eqnarray*}}
\newcommand{\eea}{\end{eqnarray*}}
\newcommand{\bean}{\begin{eqnarray}}
\newcommand{\eean}{\end{eqnarray}}
\newcommand{\eqs}[1]{Eqs. (\ref{#1})}
\newcommand{\eq}[1]{Eq. (\ref{#1})}
\newcommand{\meq}[1]{(\ref{#1})}
\newcommand{\fig}[1]{Fig. \ref{#1}}
\newcommand{\ppa}[2]{\left(\frac{\partial}{\partial #1}\right)^{#2}}

\newcommand{\ppn}[2]{\frac{\partial #1}{\partial #2}}

\newcommand{\eqn}{&=&}
\newcommand{\non}{\nonumber \\}
\newcommand{\oh}{\frac{1}{2}}

\title{Testing cosmic censorship conjecture near extremal black holes  with cosmological constants}
\author{Yuan Zhang\thanks{zhangyuan@mail.bnu.edu.cn}, Sijie Gao\thanks{Corresponding author. sijie@bnu.edu.cn}\\
Department of Physics, Beijing Normal University,\\
Beijing 100875, China}
\begin{document}
\maketitle
\begin{abstract}
It has been shown previously that an extremal Reissner-Nordstr\"om or an extremal Kerr black hole cannot be overcharged or overspun by a test particle if radiative and self-force effects are neglected.
In this paper, we consider extremal charged and rotating black holes with cosmological constants. By studying the motion of test particles, we find the following results: An extremal Reissner-Nordstr\"om anti-de Sitter (RN-AdS) black hole can be overcharged by a test particle but an extremal Reissner-Nordstr\"om de Sitter (RN-dS) black hole cannot be overcharged. We also show that both extrmal
Kerr-de-Sitter (Kerr-dS) and Kerr-anti-de-Sitter (Kerr-AdS) black holes can be   overspun by a test particle, implying a possible breakdown of the cosmic censorship conjecture. For the Kerr-AdS case, the overspinning requires that the energy of the particle be negative, a reminiscent of the Penrose process. In contrast to the extremal RN and Kerr black holes, in which cases the cosmic censorship is upheld, our results suggest some subtle relations between the cosmological constants and the comic censorship. We also discuss the effect of radiation reaction for the Kerr-dS case and find that the magnitude of energy loss due to gravitational radiation may not be enough to prevent the violation of the cosmic censorship. 

\end{abstract}

\section{Introduction}
It is generally believed that singularities appear only after the formation of black holes during gravitational collapses,  i.e., singularities exist inside black hole horizons and cannot be seen by distant observers. This is the well-known ``cosmic censorship'' conjecture first proposed by Penrose \cite{penrose}. One way of testing the cosmic censorship is to throw a test particle into an existing black hole. If the particle can pass through the horizon and change the parameters of the black hole such that the horizon disappears, then the cosmic censorship conjecture might break down. It has been shown by Wald\cite{wald72} that a test particle cannot destroy the horizon of an extremal Kerr-Newman black hole. This issue has been revisited in recent years \cite{hubeny}-\cite{semiz2}. By noticing that linear approximation had been used  in Wald's analysis, we took into account higher-order terms and find possible violation of the cosmic censorship for extremal Kerr-Newman black holes \cite{gao-zhang}.

The acceleration of our universe indicates the existence of dark energy, while the cosmological constant $\Lambda$ turns out to be a good candidate. The recent development of the AdS/CFT correspondence has given many new insights into the nature of black holes. Thus, the study of asymptotically de Sitter and anti-de Sitter black holes have become more important and realistic than ever. Previously \cite{hubeny} \cite{ted} \cite{gao-zhang}, it has been shown that the horizon of an extremal
RN or Kerr black hole cannot be destroyed by any test particles even higher order terms are considered. It is then interesting to know whether the presence of $\Lambda$ could make a difference. For charged black holes with cosmological constants, we study the extremal RN-dS and RN-AdS black holes and the conditions that the horizons can be destroyed by a test particle. We find that a particle could destroy the horizon of an extremal RN-Ads black hole, but not an extremal RN-dS black hole. In case of rotating black holes, Our analysis shows that both extremal Kerr-dS and Kerr-AdS black holes can be overspun by test particles. The interesting feature of overspinning an Kerr-AdS black hole is that the particle must possess negative energy, just like the Penrose process.

We further find that for a possible violation of the cosmic censorship, the allowed range for the particle's energy $E$ is of  order $Q\Lambda q^2$ or $L^2/M^3$, $ML\Lambda$. Since $\Lambda M^2\ll 1$ and $M\sim Q$ in the extremal case, we see that the energy of the particle must be finely tuned. We study the orbit near the horizon of the Kerr-dS black hole and estimate the energy loss due to the radiative effect. Due to the coincidence of the location of the light ring and the extremal horizon,  the radiative effect is negligible.  Such analysis has also been made by other authors for nearly extremal Kerr and RN black holes \cite{vitor-prl}-\cite{poisson}, where the radiative effect is important at least for some orbits.

\section{Degenerate horizons of RN-dS and RN-AdS black holes} \label{horizon}

A spherically symmetric black hole carrying mass $M$ and charge $Q$ and a nonvanishing cosmological constant $\Lambda$ is described by the RN-(anti-)dS metric \cite{lake}
\bean
ds^2=-f(r)dt^2+\frac{dr^2}{f(r)}+r^2 d\Omega^2\,,
\eean
where
\bean
f(r)=1-\frac{2M}{r}+\frac{Q^2}{r^2}-\frac{\Lambda}{3}r^2 \label{frlam}\,.
\eean
We first study the RN-dS metric ($\Lambda>0$). If $\Lambda$ vanishes, the metric reduces to the RN solution. In the RN case, when $M>Q$, there are two horizons determined by $f(r_i)=0$, $i=1,2$, which are called the Cauchy horizon and event horizon. For $M=Q$, the two horizons coincide with each other and $f'(r_i)=0$. When a positive $\Lambda$ is turned on, the two roots of $f(r)=0$ have negligible changes  due to the fact $\Lambda M^2\ll 1$. But with the increase of $r$, the last term in \eq{frlam} finally dominates and a third horizon $r_3$ appears. This horizon is called the cosmological horizon.

\begin{figure}[htmb]
\centering \scalebox{0.6} {\includegraphics{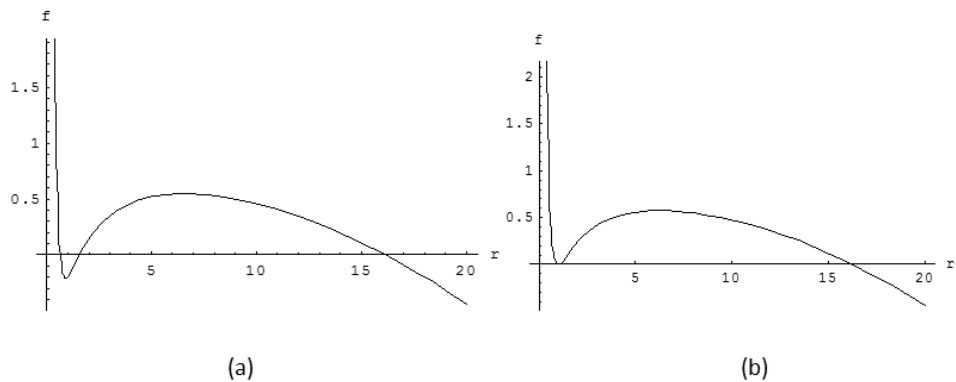}}
\caption{Function $f(r)$ of  RN-dS black holes. (a): An ordinary RN-dS black hole with three horizons. (b): An extremal black hole containing a degenerate horizon}  \label{fr3}
\end{figure}

\fig{fr3} (a) shows the function $f(r)$ of a RN-dS black hole with all the three horizons. If $r_1=r_2$, the two horizons coincide and we have $f'(r_1)=f(r_1)=0$. This means that the black hole possesses a degenerate horizon which corresponds to a zero temperature according to black hole thermodynamics. In this case, $f(r)$ is depicted in \fig{fr3} (b).  There is also a possibility that the event horizon coincides with the cosmological horizon, i.e., $r_2=r_3$. We shall not discuss this case because an extremal RN-dS black hole usually refers to the degenerate horizon depicted in \fig{fr3} (b) and the cosmological horizon is not relevant to the discussion of the cosmic censorship.

Now we investigate under what conditions an extremal black hole can exist. Denote $r_h\equiv r_1=r_2$, which satisfies
\bean
f(r_h)=0
\eean
and then leads to the relation
\bean
M=\frac{3Q^2+3r_h^2-\Lambda r_h^4}{6r_h}\label{mfqq}\,.
\eean
An extremal black hole requires that  $f'(r_h)=0$. It is not difficult to verify that this condition is equivalent to
\bean
\ppn{M}{r_h}=0\,.
\eean
Thus, the degenerate event horizon is located at\footnote{This solution is valid for either $\Lambda>0$ or $\Lambda<0$.  There is another solution $r_h=\sqrt{\frac{1+\sqrt{1-4Q^2\Lambda}}{2\Lambda}}$ for $\Lambda>0$, which corresponds to the cosmological horizon as we discussed above. }
\bean
r_h= \sqrt{\frac{1-\sqrt{1-4Q^2\Lambda}}{2\Lambda}} \label{rhfr} \,.
\eean
Here and through this paper, we assume $|\Lambda Q^2|\ll 1$. This means that $\Lambda$ can be treated as a perturbation to the RN solution around the region $r\sim r_h$ .  Using \eqs{mfqq} and \meq{rhfr}, one  can write  $M$ as a function of $Q$:
\bean
M=M(Q)=\frac{1+4Q^2\Lambda-\sqrt{1-4Q^2\Lambda}}{3\sqrt{2}\Lambda
\sqrt{\frac{1-\sqrt{1-4Q^2\Lambda}} {\Lambda} } } \,.
\label{mfq}
\eean
If the black hole parameters satisfy this equation, we have an extremal black hole as shown in \fig{fr3} (b). If we decrease the mass $M$ by a small amount $\Delta M$ while keeping $Q$ fixed, \eq{frlam} clearly shows that $f(r)$ will increase everywhere and the degenerate root will disappear, implying the disappearance of the extremal horizon. Therefore, \eq{mfq} gives the minimum mass for the existence of a horizon.

For a RN-AdS black hole $\Lambda<0$, all the above arguments and results  are valid except that there is no cosmological horizon.

\section{Overcharging RN-(anti-)dS black holes}

Assuming a test particle is moving towards a RN-(anti-)de Sitter black hole in the radial direction with four-velocity
\begin{equation}
u^{a}=\dot{t}\left(\frac{\partial}{\partial t}\right)^{a}+\dot{r}\left(\frac{\partial}{\partial r}\right)^{a} \,.
\end{equation}
The four-potential of the vacuum electromagnetic field is
\begin{equation}
A_{a}=A_{t}dt_{a}+A_{\phi}d\phi_{a} \,.
\end{equation}
The energy of the particle can be defined through the killing vector field $(\partial/\partial t)^{a}$
\begin{equation}
E=-t^{a} (m u_{a}+q A_{a})=-m g_{tt}\dot{t}-q A_{t}=m\dot{t}f(r)+\frac{qQ}{r} \label{etam}\,,
\end{equation}
where $\dot{t}>0$ because $u^a$ is future directed. Since $f(r)>0$ outside the black hole horizon $r=r_h$, we obtain from \eq{etam}
\bean
E>\frac{qQ}{r_h}\equiv E_{min} \label{em1} \,.
\eean
This is the condition that a particle can be captured by the black hole. To overcharge the black hole, we need another constraint on the parameters of the particle and black hole.

As argued at the end of Section \ref{horizon},  $M(Q)$ defined by \eq{mfq}  is the critical mass for a given $Q$. If the black hole mass is smaller than $M(Q)$, there is no event horizon. In attempt to destroy the horizon of an existing extremal black hole, a particle with energy $E$ and charge $q$ must satisfy
\bean
M(Q)+E<M(Q+q) \,.
\eean
So the upper limit for $E$ is
\bean
E<M(Q+q)-M(Q) \equiv E_{max} \label{em2} \,.
\eean
To destroy the horizon, the particle must satisfy both inequalities \meq{em1} and \meq{em2}, which means
\bean
E_{max}-E_{min}>0 \label{emm} \,.
\eean
Using \eqs{rhfr} and \meq{mfq}, we can express the left-hand side of \eq{emm} as a function of $(Q,q,\Lambda)$. Since $q$ and $\Lambda$ are small values, by Taylor expansion, we have the leading term
\bean
E_{max}-E_{min}=-\frac{|Q|}{2}\Lambda q^2+O(\Lambda^2 q^2)+O(\Lambda q^3)+... \label{emle}\,.
\eean
This shows that \eq{emm} holds only for $\Lambda <0$. There is no violation of the cosmic censorship for extremal RN-dS black holes.

For confirmation, we take $Q=1$, $\Lambda=-10^{-4}$ and $q=1.3\times 10^{-3}$. The mass of the extremal RN-Ads black hole is determined by \eq{mfq}.  Substituting these data into \eq{em1}, we find
\bean
E_{min}\approx 1.30\times 10^{-3} \,.
\eean
Using \eq{em2} to compute $E_{max}$, we have
\bean
\Delta E=E_{max}-E_{min}=8.45\times 10^{-11}  \,.
\eean
Therefore, as long as $E$ is chosen within this range, we can always have solutions satisfying both \eq{em1} and \eq{em2}, meaning that the extremal RN-AdS black holes could be destroyed.
Note that $\Delta E$ is of order $Q\Lambda q^2$. For a particle to attain an energy within this range, a fine tuning is required.

\section{Overspinning   Kerr-dS and Kerr-AdS black holes}

Now we turn attention to rotating black holes with cosmological constants, i.e., Kerr-dS and Kerr-Ads solutions. Such a spacetime is described by the metric \cite{kerrds}
\begin{equation}
ds^{2}=-\frac{\Delta_{r}}{\chi^{2}\rho^{2}}(dt-a \sin^{2}\theta d\phi)^{2}+\frac{\rho^{2}}{\Delta_{r}}dr^{2}
+\frac{\Delta_{\theta}\sin^{2}\theta}{\chi^{2}\rho^{2}}(adt-(r^{2}+a^{2})d\phi)^{2}+\frac{\rho^{2}}{\Delta_{\theta}}d\theta^{2}
\label{kmetric}
\end{equation}
with
\begin{equation}
\begin{split}
& \chi=1+\frac{a^{2}\Lambda}{3}\\
& \rho^{2}=r^{2}+a^{2}\cos^{2}\theta\\
& \Delta_{r}=(1-\frac{\Lambda}{3} r^{2})(r^{2}+a^{2})-2Mr\\
& \Delta_{\theta}=1+\frac{a^{2}\Lambda}{3} \cos^{2}\theta
\end{split}
\end{equation}
where $M$ is the mass of the black hole and $a=J/M$ is the specific angular momentum.

A horizon is located at
\bean
\Delta_r(r_h)=0 \label{hork}\,.
\eean

By the same argument as given in \cite{gao-zhang}, we find the condition for the particle to be captured by the black hole is
\begin{equation}
E\geq\frac{-g_{t\phi}L}{g_{\phi\phi}}\Big|_{r=r_h} \,.
\end{equation}
By direct substitution, one finds
\begin{equation}
E\geq\frac{aL}{r_h^{2}+a^{2}}\equiv E_{min}\label{ebrh} \,.
\end{equation}

In order to find the condition to overspin the black hole, we first use
\eq{hork} to express the mass in terms of the horizon radius
\bean
M=\frac{a^2}{2r_h}+\frac{r_h}{2}-\frac{1}{6}a^2\Lambda r_h -\frac{\Lambda}{6}r_h^3\label{mfr}\,.
\eean
An extremal black hole satisfies $f'(r_h)=0$, which is equivalent to
\bean
\ppn{M}{r_h}=\oh-\frac{a^2}{2r_h^2}-\frac{\Lambda}{6}a^2-\frac{\Lambda}{2}r_h^2=0\label{mpr}\,.
\eean
By solving \eq{mpr}, one has the solution of the horizon radius
\bean
r_h=\sqrt{\frac{3-\Lambda a^2-\sqrt{\Lambda^2a^4-42\Lambda a^2+9}}{6\Lambda}}\label{ral}\,.
\eean
\eq{mfr} together with \eq{ral} defines a function
\bean
M=f_M(a) \label{fma}\,.
\eean
A black hole is extremal if and only if this relation is satisfied. For a fixed $a$, \eq{fma} also gives the minimum mass for the existence of a horizon.  Consider an extremal black hole with $M$ and $a$. A particle with $E$ and $L$ is dropped into the black hole. The final state of the black hole is described by
\bean
M'\eqn M+E \\
a'\eqn \frac{Ma+L}{M+E}\,.
\eean
Thus, the condition for a possible violation of the cosmic censorship is given by
\bean
M'<f_M(a')\,,
\eean
i.e.,
\bean
M+E<f_M\left(\frac{Ma+L}{M+E}\right) \label{mef}
\eean
To proceed, we need to simplify \eq{fma}. By expanding $M$ to the first order of $\Lambda$, we find
\bean
f_M(a)\approx |a|-\frac{1}{3}|a|^3 \Lambda \label{expm}\,,
\eean
Without loss of generality, we shall assume $a>0$ in the following calculation (Obviously, the sign of $a$ should not change the nature of our analysis).
Then \eq{mef} is approximated as
\bean
M+E<\left(\frac{M a+L}{M+E}\right)-\frac{\Lambda}{3}\left(\frac{Ma+L}{M+E}\right)^3
\eean
Define
\bean
W=(M+E)^2\,,
\eean
which satisfies
\bean
W^2-(Ma+L)W+\frac{\Lambda}{3}(Ma+L)^3<0\,.
\eean
The solution of this inequality is
\bean
z_1<W<z_2\,,
\eean
where
\bean
z_2=\frac{1}{6}(3L+3aM+\sqrt{(-3L-3aM)^2-12(L^3\Lambda+3aL^2M\Lambda+3a^2LM^2\Lambda+a^3M^3\Lambda})
\label{z2}
\eean
and $z_1<0$.

Similarly, \eq{ebrh} can be rewritten as
\bean
W\equiv (E+M)^2\geq \left(\frac{aL}{r_h^2+a^2}+M\right)^2\equiv z_3 \label{z3}\,.
\eean
Then the necessary and sufficient condition for the particle to destroy the horizon is
\bean
z_2-z_3>0 \label{z23b}\,.
\eean
Using \eq{expm} to replace $M$ in \eqs{z2} and \meq{z3}, \eq{ral} to replace $r_h$, and then Taylor expanding $z_2-z_3$ around $\Lambda=L=0$, we finally obtain
\bean
z_2-z_3=-\frac{L^2}{4a^2}+\frac{a^2}{3}L\Lambda+O(L\Lambda^2)+... \label{z23}
\eean
This corresponds to
\bean
\Delta E\equiv E_{max}-E_{min}\sim L^2/M^3\sim ML\Lambda\,,
\eean
which means that the allowed energy window for the particle is very narrow.

Using \eq{z23},  the solution of \eq{z23b} is found to be
\bean
0<L<\frac{4}{3}\Lambda a^4 \label{lfra}\,,
\eean
or
\bean
0>L>\frac{4}{3}\Lambda a^4 \label{lfrb}\,.
\eean
Clearly, \eq{lfra} requires $\Lambda>0$ and \eq{lfrb} requires $\Lambda<0$. In the following, we shall discuss the two cases respectively. \\

\textbf{1. Kerr-de Sitter solution ($\Lambda>0$)}

For $\Lambda>0$, we may choose a small but positive $L$ to satisfy the inequality \meq{lfra}, leading to a possible violation of the cosmic censorship. One might suspect that the terms we dropped above could cause some serious error and then invalidate our conclusion.  For clarification,  we construct a numerical solution as follows.  We choose
\bean
a=1,\ \ \ \Lambda=10^{-4}, \ \ \ L=10^{-4}
\eean
so that the inequality \meq{lfra} holds. \eq{fma} gives the mass of the existing extremal Kerr-dS black hole as
\bean
M=0.999967\,.
\eean
We use \eqs{ebrh} and  \meq{ral} to find the lower limit of $E$
\bean
E_{min}=4.99967\times 10^{-5}\,.
\eean
Choose the energy of the particle to be
\bean
E=E_{min}+10^{-10}
\eean
and we find
\bean
f_M\left(\frac{Ma+L}{M+E}\right)-(M+E)=6.336\times 10^{-10}>0 \,.
\eean
Therefore, we have found a solution such that \eqs{mef} and \meq{ebrh} both hold. Note that our numerical calculation only involves original formulas without any approximation.

\textbf{2. Kerr-anti-de Sitter solution ($\Lambda<0$)--Penrose process}

\eq{lfrb} shows that both $L$ and $\Lambda$ are negative. As a consequence, the energy $E$ of the particle must be negative too. The reason is as follows.

In the absence of $\Lambda$, i.e., the Kerr solution, the overspinning condition \meq{mef} simply reduces to
\bean
M+E<\frac{Ma+L}{M+E}\,.
\eean
With the extremal condition $M=a$, we have
\bean
E<\frac{L}{2M}\,.
\eean
Obviously,  a negative angular momentum $L$ implies a negative energy $E$ in the Kerr case.
On the other hand, one see immediately that \eq{ebrh} reduces to
\bean
E\geq \frac{L}{2M}\,,
\eean
because $M=a=r_h$ in the extreme Kerr solution. Therefore, no overspinning actually occurs in the absence of $\Lambda$.

Now consider the case $\Lambda<0$ and $L<0$. With the help of \eq{expm}, one can expand \eq{mef} around $\Lambda=0$ and $L=0$:
\bea
s&\equiv& f_M\left(\frac{Ma+L}{M+E}\right)-(M+E) \non
\eqn -a-E+\frac{a^2}{a+E}+\frac{L}{a+E}+\left(\frac{a^3}{3}-\frac{a^6}{3(a+E)^3}+\frac{a^5}{3(a+E)^2}
-\frac{a^4}{3(a+E)}\right)\Lambda+O(\Lambda L) \,.
\eea
Since $E\ll a$, $s$ can be expanded as
\bean
s=\frac{L}{a}+\left(-2-\frac{L}{a^2}+\frac{2a^2\Lambda}{3}\right)E\,.
\eean
Noting that $|L|\ll a$ and $\Lambda a^2\ll 1$, we see that $s>0$ implies $E<0$. Therefore, to overspin an extremal Kerr-AdS black hole, the energy of the particle must be negative. This is a reminiscent of the Penrose process \cite{waldbook}. In the Kerr-AdS spacetime, an ergo sphere exists outside the black hole horizon. Thus, a particle with $E<0$ can be made within the ergo sphere. In the Kerr spacetime, it is well-known that energy can be extracted from the black hole via the Penrose process. However, in the Kerr-AdS spacetime, our derivation suggests that a particle with finely tuned negative energy can enter the horizon and then make it disappear. This is an important difference caused by the negative cosmological constant in the Penrose process. This result is consistent with that obtained by Cardoso, Dias and Yoshida \cite{super} who showed that small Kerr-AdS black holes are unstable against scalar perturbation, via a superradiant mechanism.

To verify the conclusion numerically, we choose
\bean
a=1, \ \Lambda=-10^{-4},\ L=10^{-4}\,.
\eean
The black hole mass determined by \eq{fma} is given by
\bean
M=1.000033\,.
\eean
Similarly to the case $\Lambda>0$, we find that the lower limit of the particle's energy is
\bean
E_{min}=-5.00033\times 10^{-4}\,.
\eean
Choose the energy of the particle to be slightly bigger than the lower limit
\bean
E=E_{min}+10^{-10}\,.
\eean
We then find
\bean
f_M\left(\frac{Ma+L}{M+E}\right)-(M+E)=6.330\times 10^{-10}>0\,.
\eean
Therefore, we have found a particle with negative energy which satisfies both \eqs{mef} and \meq{ebrh}.


\section{Discussion of radiative effects in the Kerr-dS case} \label{sec-rad}
In the above analysis, we treated the spacetime as a fixed background and neglected radiation reaction from the particle. We have mentioned that the allowed energy window is quite small for the Kerr-dS case. So if the gravitational radiation of the particle is of the same order, the overspinning might be prevented. Recently, Barausse, Cardoso and Khanna \cite{vitor-prl, vitor-prd} showed that the effect of radiation reaction and self-force could help prevent the formation of naked singularities. A key point in their analysis is that when possible violation of the cosmic censorship occurs, the particle always follows an almost circular orbit such that the particle may radiate significant energy before it reaches the horizon. Following the same spirit, we analyze the orbits near the horizon and find that the particle will make tremendously large number of circles near the horizon and cause energy loss that is sufficient to break down the argument in the previous section.

We consider the the Kerr-dS case. Substituting \eq{ral} into \eq{ebrh} and expand at $\Lambda=0$, we find
\bean
E_{min}=\frac{L}{2a}-\frac{1}{3}aL\Lambda\,,
\eean

Expanding \eq{z2} around $\Lambda=0$ gives
\bean
E_{max}=\frac{L}{2a}-\frac{L^2}{8a^3}-\frac{aL}{6}\Lambda\,.
\eean
Then
\bean
E_{max}-E_{min}>0
\eean
yields
\bean
0<L<\frac{4}{3}a^4\Lambda\,,
\eean
which is the same as \eq{lfra}. We can take
\bean
L=a^4\Lambda \label{la4}
\eean
such that
\bean
E_{min}\eqn \frac{L}{2a}-\frac{1}{3}a^5\Lambda^2  \label{em1} \\
E_{max}\eqn \frac{L}{2a}-\frac{7}{24}a^5\Lambda^2  \label{em2}
\eean
Then the allowed energy is written in the form
\bean
E=\frac{L}{2a}-\beta a^5\Lambda^2 \label{ela}
\eean
with
\bean
\frac{7}{24}<\beta<\frac{1}{3}\,.
\eean

The standard calculation shows that the equations of motion for a particle with mass $m$ on the $\theta=\frac{\pi}{2}$ in the Kerr-dS spacetime are given by
\bean
\dot r^2\eqn \frac{E^2 g_{\phi\phi}+2E g_{t\phi} L-g_{t\phi}^2 m^2+g_{tt}(L^2+g_{\phi\phi} m^2)}{g_{rr}(g_{t\phi}^2-g_{\phi\phi}g_{tt})m^2}\,,\non
\dot\phi\eqn-\frac{Eg_{t\phi}+g_{tt}L}{(g_{t\phi}^2-g_{\phi\phi}g_{tt})m}\,. \label{fdt}
\eean
Assuming $E\gg m$\cite{ted, vitor-prl, vitor-prd}, we have
\bean
\dot r^2\eqn \frac{E^2 g_{\phi\phi}+2E g_{t\phi} L+g_{tt}L^2}{g_{rr}(g_{t\phi}^2-g_{\phi\phi}g_{tt})m^2}\equiv V(r)\,.
\eean
Near the horizon $r=r_h$, $V(r)$ may be written as
\bean
V(r)\approx V(r_h)+V'(r_h)(r-r_h) \label{vrv}\,.
\eean

Substituting the metric \meq{kmetric}, where $M$ satisfies \eq{expm}, and using \eqs{la4} and \meq{ela}, then expanding $V(r)$ around $\Lambda=0$, we find   
\bean
V(r_h)\approx \frac{4a^{10}\Lambda^4}{9m^2}(1-3\beta)^2 \,,
\eean
and
\bean
V'(r_h)=\frac{4a^7(1-3\beta)\Lambda^3}{3m^2}>0 \,.
\eean

The liner approximation \meq{vrv} is valid within the range
\bean
\Delta r=r-r_h\sim \frac{V'(r_h)}{V''(r_h)}\sim (1-3\beta)\Lambda a^3\,.
\eean

On the other hand, it follows from \eq{fdt} that
\bean
\dot\phi\sim\frac{a^3\Lambda}{m(r-r_h)}\sim \frac{1}{m}
\eean
at $r_h$.
Therefore,
\bean
\frac{d\phi}{dr}\approx \frac{1}{m\sqrt{V(r_h)+V'(r_h)(r-r_h)}}\,.
\eean

Finally, the number of cycles made by the particle near the horzion is given by
\bean
N\eqn \int_{r_h}^{r_h+\Delta r} \frac{1}{m\sqrt{V(r_h)+V'(r_h)(r-r_h)}} dr
\sim \frac{1}{\Lambda a^2}\,.
\eean
For a black hole with 100 solar mass, one can show $\Lambda a^2\sim 10^{-42}$. So $N$ is an extremally large number.

According to \cite{vitor-prl}, the gravitational-wave loss of energy can be estimated by
\bean
E_{rad}\sim N (r_0-r_h)E^2/a^2\,,
\eean
where $r_0$ is the radius of the circular phonon orbit and we have divided the formula in \cite{vitor-prl} by $a^2$ to make the dimension consistent with our convention.  However, as shown in Appendix \ref{app-1}, $r_0=r_h$ for the extremal Kerr-dS black hole, which makes $E_{rad}$ vanish. Unlike the results for nearly extremal Kerr black holes \cite{vitor-prl}-\cite{vitor-prd}, which suggest that the radiative effects can be neglected for some trajectories giving rise to naked singularities, we show that the radiative effects are negligible for all trajectories that lead to possible violation of the cosmic censorship.

\section{Conclusions}

We have extended the test of the cosmic censorship to black holes with cosmological constants. Previous works showed that extremal RN and Kerr black holes cannot serve as counterexamples of the cosmic censorship even if higher order terms are taken into account. In the presence of $\Lambda$, we derived the condition that a test particle can be absorbed by an extremal black hole and the condition that the black hole horizon might disappear after the absorbtion. For charged black holes, we show that an extremal RN-AdS black hole can be overcharged, implying possible violation of the cosmic censorship. In contrast, an extremal RN-dS black hole cannot be overcharged. For rotating black holes, both extremal Kerr-dS and Kerr-AdS black holes can be overspun. In particular, to overspin the Kerr-AdS black hole, the energy of the particle must be negative, i.e., the Penrose process is involved.

Similarly to the discussion in the absence of $\Lambda$, we have found that the allowed parameter range for the test particle is very narrow. By studying the orbits near the  horizon of an extremal Kerr-dS black hole and using the light ring argument, we found that the  radiative effect is not enough to prevent the violation of the cosmic censorship. As shown in \cite{vitor-prl}-\cite{vitor-prd}, the self-force effect is important for nearly extremal black holes. Although we have not discussed the self-force effect in this paper, our results suggest some interesting regime in which such effect should be studied.

\section*{Acknowledgements}
We thank an anonymous referee for helpful comments which leads to the discussion on the radiative effect in the revision. This research was supported by NSFC Grants No. 11235003, 11375026 and NCET-12-0054.

\appendix
\section{Circular photon orbit} \label{app-1}
We  calculate the circular photon orbit (light ring).
Consider the orbit of a photon on the $\theta=\frac{\pi}{2}$ plane in the Kerr-dS spacetime. The tangent vector $k^a$ to an orbit parameterized by $\lambda$ is written as
\bean
k^a=\dot t \ppa{t}{a}+\dot r \ppa{r}{a}+\dot \phi\ppa{\phi}{a}\,.
\eean
The conservation of energy and angular momentum leads to
\bean
E\eqn -g_{ab}k^a\ppa{t}{a}\,, \\
L\eqn g_{ab}k^a\ppa{\phi}{a}\,.
\eean
Together with the constraint
\bean
g_{ab} k^a k^b=0
\eean
and the metric \eq{kmetric}, we find the radial motion for the photon is given by
\bean
\dot r^2=\frac{1}{27r^3}V(r)\,,
\eean
where
\bean
V(r)\eqn   9 E^2 \left(6 M+(\Lambda +3) r \left(r^2+1\right)\right)-6 E (\Lambda +3) L \left(6 M+\Lambda  r \left(r^2+1\right)\right)\non
&+&(\Lambda +3)^2 L^2 \left(6 M+r \left(\Lambda +\Lambda  r^2-3\right)\right) \,. \label{vrr}
\eean
Without loss of generality, we have taken $a=1$ in \eq{vrr}. The light ring refers to the circular orbit at $r=r_0$ satisfying
\bean
V'(r_0)=V(r_0)=0\,.
\eean
First,  $V'(r_0)=0$ yields
\bean
r_0=\frac{\sqrt{-27E^2+27 L^2-9E^2\Lambda+18EL\Lambda+9L^2\Lambda+6EL\Lambda^2-3L^2\Lambda^2-L^2\Lambda^3}}
{\sqrt{3}\sqrt{27E^2+9E^2\Lambda-18EL\Lambda+9L^2\Lambda-6EL\Lambda^2+6L^2\Lambda^2
+L^2\Lambda^3}}\,.
\label{rnaut}
\eean

Our purpose is to calculate $r_0$. So without loss of generality, we choose $E=1$. By substituting \eq{fma} into \eq{vrr}, we can solve $V(r_0)=0$ and \eq{rnaut} for $L$ and $r_0$. It is difficult to find the analytical solutions. However, we shall show that $r_0$ is coincident with the horizon radius $r_h$. For this purpose, we assume
\bean
r_0=r_h\,,
\eean
where $r_h$ is given by \eq{ral}. Thus, $L$ can be solved as
\bean
L \eqn\frac{\sqrt{32 \Lambda ^4-768 \Lambda ^3-96 \sqrt{\Lambda ^2-42 \Lambda +9} \Lambda ^2+288 \Lambda ^2+32 \sqrt{\Lambda ^2-42 \Lambda +9} \Lambda ^3}-6 \Lambda }{2 \left(2 \Lambda ^3+6 \Lambda ^2\right)}\non
&+&\frac{14 \Lambda ^2+2 \sqrt{\Lambda ^2-42 \Lambda +9} \Lambda}{2 \left(2 \Lambda ^3+6 \Lambda ^2\right)}\,.
\eean
Substituting $L$ into \eq{vrr}, we obtain
\bean
V(r_h)&\propto&
\sqrt{2} \Lambda ^2+\sqrt{2} \Lambda\sqrt{\Lambda ^2-42 \Lambda +9}-\left(\sqrt{\Lambda ^2-42 \Lambda +9}-3\right)\times \non
&& \left(3 \sqrt{2}-\sqrt{\Lambda ^2+\left(\sqrt{\Lambda ^2-42 \Lambda +9}-24\right) \Lambda -3 \sqrt{\Lambda ^2-42 \Lambda +9}+9}\right)\non
&+&\left(\sqrt{\Lambda ^2+\left(\sqrt{\Lambda ^2-42 \Lambda +9}-24\right) \Lambda -3 \sqrt{\Lambda ^2-42 \Lambda +9}+9}-24 \sqrt{2}\right) \Lambda \,.\non
\
\eean
By some algebraic manipulation, we find
\bean
V(r_h)=0 \,.
\eean
for all $\Lambda$. Therefore, $r_0$ is equal to $r_h$.

\end{document}